# Optically assisted diffusophoretic tweezers using resonant plasmonic bowtie nano-antennas

THEODORE ANYIKA,[1,3] IKJUN HONG,[1,3] AND JUSTUS NDUKAIFE[1,2,3,4*]

[1]*Department of Electrical and Computer Engineering, Vanderbilt University, Nashville, TN, USA*
[2]*Department of Mechanical Engineering, Vanderbilt University, Nashville, TN, USA*
[3]*Vanderbilt Institute of Nanoscale Science and Engineering*
[4]*Vanderbilt Center for Extracellular Vesicles Research*
*justus.ndukaife@vanderbilt.edu*

**Abstract:**

Plasmonic antennas, leveraging localized surface plasmon resonance (LSPR), hold significant promise for efficiently trapping nanoscale particles at low power levels. However, their effectiveness is hindered by photothermal effects in metallic nanoparticles, leading to repulsive thermophoretic forces. To address this limitation, we propose a novel hybrid approach that combines depletion attraction and photothermal effects inherent in plasmonic structures, capitalizing on thermally induced concentration gradients. Through the thermophoretic depletion of polyethylene glycol (PEG) molecules around plasmonic hotspots, we create sharp concentration gradients, enabling precise localization of nanoscopic particles through a synergistic effect with diffusophoretic forces. In our experiments, we successfully demonstrate the trapping and dynamic 2D manipulation of $100\ nm$ polystyrene beads, showcasing the platform's potential for colloidal assembly at the nanoscale. Remarkably, this method maintains highly stable trapping performance even at an incredibly low $2.5\ mW$ laser power, rendering it particularly appealing for applications involving biological species. Our study introduces a promising avenue for the precise and efficient manipulation of sub-nanoscale particles, with wide-ranging implications in nanotechnology, biophysics, and nanomedicine. This research opens up new opportunities for advancing nanoscale particle studies and applications, ushering in a new era of nanoscale manipulation techniques.

## 1. Introduction

Plasmonic nanostructures have garnered considerable attention as a solution for efficient low-power trapping of nanoscale particles, leveraging their localized surface plasmon resonance (LSPR), which enables the generation of high field gradients well beyond the diffraction limit [1–4]. Notably, LSPR resonances lead to a significant increase in the scattering and absorption cross-sections, resulting in significant heat generation around metallic nanoparticles [3,4]. These photothermal effects in metallic particles have been leveraged for applications such as photothermal therapy for cancer treatment [5,6]. However, temperature gradients formed at the interface between the metal and fluid environment induce various thermal effects, such as thermophoresis, thermo-osmosis, Marangoni convection, and thermoplasmic convection, which typically decrease the stability of trapped particles in the vicinity of plasmonic antennas [7–9].

Although plasmonic tweezers have shown promising results, their effectiveness are often constrained by these thermoplasmic effects. For instance, Gargiulo *et al.* [7] demonstrated that, thermophoretic and thermo-osmotic effects impose a limiting inter-particle distance of approximately 300 $nm$ in the 2D optical assembly of metallic nanoparticles. However, these thermal gradients can be harnessed to induce thermoelectric [10–12] or depletion [13–15] forces which enable precise localization of nanoscopic objects. Thermoelectric trapping necessitates the use of an ionic surfactant, such as cetyltrimethylammonium chloride (CTAC). This surfactant forms positively charged micelles with a high Soret coefficient, accompanied by negatively charged chlorine ions exhibiting a lower Soret coefficient, which results in the formation of a spatially-distributed thermoelectric field due to the difference in the Soret coefficients. The use of surfactants for thermoelectric trapping however limits its applicability to bio species as surfactants typically disrupt lipid structures [16]. Thermophoretic trapping using depletion attraction forces relies on the depletion of polymers like polyethylene glycol (PEG) away from the hot region, resulting in a concentration gradient $\nabla c$. This gives rise to a diffusophoretic force $F_d = -3\pi\eta a D S_T^* \nabla T$ on particles in the depleted region, where $a$ is the particle diameter, $\eta$ is the viscosity of the solution, $D$ is the Brownian diffusion coefficient of the particle, $S_T^*$ is the effective Soret coefficient of the particle which is proportional to the PEG concentration ($c_{PEG}$), and $\nabla T$ is the temperature gradient. Several works utilizing diffusophoresis rely on infrared laser absorption by water and subsequent heating of water to generate thermal gradients for particle aggregation [13,14,17]. To the best of our knowledge, single-particle trapping and dynamic manipulation using thermally induced PEG concentration gradients has not been demonstrated.

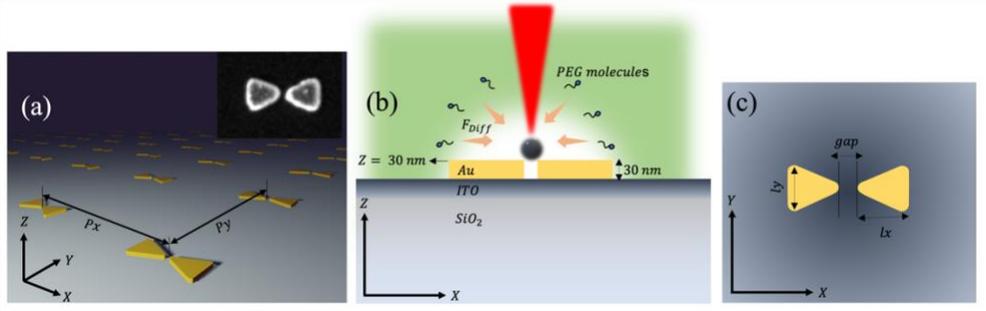

**Fig. 1.** Schematic design and working principle. (a) Schematic depiction of the bowtie array with the inset showing the scanning electron micrograph of a fabricated bowtie antenna. (b) Working principle depicting the thermophoretic depletion of PEG molecules upon laser illumination and the attractive diffusophoretic force in the XZ plane. (c) Schematic design of the bowtie antennas in the XY plane.

In this study, we introduce a novel hybrid approach that synergistically leverages the photothermal effects of plasmonic bowtie antennas to induce a highly localized diffusophoretic force, leveraging thermally induced concentration gradients. Our platform leverages the depletion attraction force resulting from thermophoretic depletion of PEG molecules surrounding a plasmonic hotspot, which overcomes the effects of repulsive thermophoresis, to achieve stable trapping of 100 $nm$ polystyrene beads. By effectively combining plasmonic heating effect with the thermally induced depletion-attraction force, we successfully demonstrate low-power trapping of nanoscale objects, with less than 3 $mW$ of power.

Additionally, we demonstrate dynamic 2D manipulation of 100 $nm$ polystyrene particles, showing the great potential of this thermoplasmonic platform.

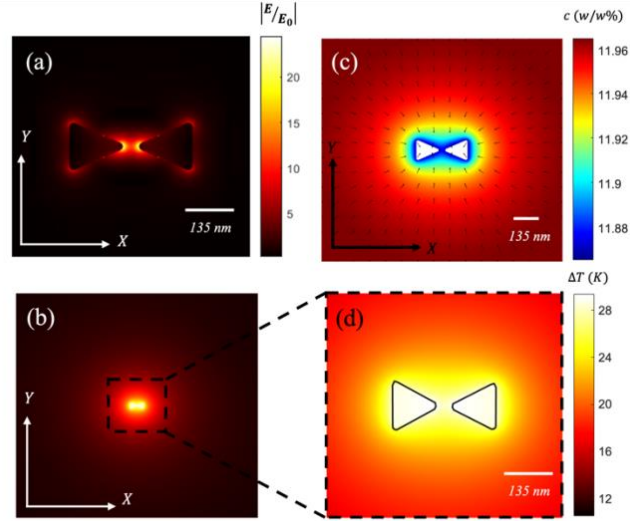

**Fig. 2.** Finite element method (FEM) simulations for resonant bowtie antennas for a 973 $nm$ laser illumination. (a) Electric field enhancement simulation in the XY plane showing a field enhancement up to 27 at Z = 15 $nm$. The incident is polarized along the X axis. (b) FEM thermal simulation in the XY plane at Z = 15 $nm$ for a 2.5 $mW$ $laser$, showing the highly localized thermal field with (d) showing the close-up thermal profile of the bowtie antennas. (c) Corresponding mass transport simulation in the XY plane at Z = 15 $nm$ for a 12% PEG solution showing the PEG concentration profile (% weight) corresponding to the thermal profile in (d). The black arrows in (c) show the direction of the diffusophoretic force.

## 2. Results

To realize efficient photothermal conversion with highly localized thermal gradients for single-particle localization, we designed plasmonic bowtie antennas resonant near $973nm$ as shown in figures 1(a-c), with $lx = 135\ nm$, $ly = 127\ nm$, $gap = 40\ nm$, $Px = Py = 1\mu m$, and a thickness of 30 $nm$. Next, we performed electromagnetic Finite Element Method (FEM) simulations using COMSOL Multiphysics simulation software to explore the photothermal conversion phenomenon in bowtie antennas placed on a glass substrate coated with a 15 $nm$ thick ITO layer. By solving the wave equation $\nabla \times \nabla \times \mathbf{E} - k_0^2 \varepsilon(r)\mathbf{E} = 0$, for an x-polarized 973 $nm$ incident plane wave, we obtained the local electric field $\mathbf{E}$ around the bowtie antennas. Here, $k_0$ represents the wave vector, and $\varepsilon(r)$ is the complex permittivity. The results show a 27 times enhancement of the local electric field $|\mathbf{E}|$, defined as $\left|E/E_0\right|$ shown in figure 2(a), where $|E_0|$ is the magnitude of the incident electric field. To characterize the photothermal heating of the bowtie antennas, we computed the heat source density $q(r) = 1/Re(\mathbf{J} \cdot \mathbf{E})$, where **J** denotes the current density induced in the plasmonic antennas. Using the heat source density, we performed FEM thermal simulations to obtain the temperature rise $T$ due to the photothermal heating as shown in figures 2(b) and (d). Further details regarding the thermal simulation are provided in the supplementary information.

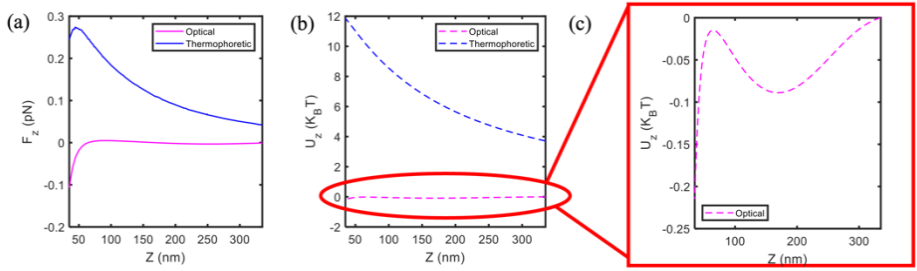

**Fig. 3.** Comparative analysis for optical and thermophoretic forces. (a) Axial forces due to optical and thermophoretic effects on a $100\ nm$ polystyrene bead under X-polarized $2.5\ mW$ laser excitation at $973\ nm$. The optical force is evaluated through the Maxwell's stress tensor method, with the bottom edge of the particle positioned along the Z-axis ranging from $Z = 35\ nm$ ($5\ nm$ from the bowtie surface) to $Z = 385\ nm$. Concurrently, the thermophoretic force is computed for the same coordinates. (b) Illustration of the optical and thermophoretic potentials, where the optical potential exhibits a shallow attractive nature, while the thermophoretic potential is significantly larger and repulsive. (d) Optical potential corresponding to the plot in (b).

Figure 2(d) shows a temperature rise of up to $29\ K$ for a $2.5\ mW$ incident laser. Thermal simulations for $4\ mW$ and $6\ mW$ incident light fields indicate a temperature rise of up to $70\ K$ (for $6\ mW$) as shown in supplementary figure 1. These findings demonstrate the efficient optothermal conversion achieved in this system. The mass transport of PEG molecules in the presence of thermal gradients is governed by the phenomenological equation, $j_{PEG} = cS_T^{PEG} D_{PEG} \nabla T - \nabla c D_{PEG}$, where $j_{PEG}$ is the flux, $c$ is the PEG concentration, $D_{PEG}$ is the Brownian diffusion coefficient of the PEG molecules, $S_T^{PEG}$ is the Soret coefficient of the PEG molecules and $\nabla T$ is the local temperature gradient. The first term on the right-hand side of the equation accounts for the migration of the PEG molecules away from the hot region due to thermophoresis, while the second term provides a counter effect due to diffusion. We have ignored the effects of thermoplasmic convection due to relatively large chamber height of $120\ \mu m$ considered here. The steady state PEG concentration profile was then computed solving the steady state mass transport equation as detailed in the supplementary information. The result in figure 2(c) shows the concentration profile for the resonant bowtie antennas illuminated using a $2.5\ mW$ laser in a $12\%$ PEG solution. This shows a $2.5\ \%$ depletion of PEG molecules from the vicinity of the bowtie antennas. $100\ nm$ polystyrene beads in the presence of large temperature gradients experience strong positive thermophoresis proportional to the Soret coefficient, which repels the particles towards cooler regions [18,19]**.** However, the presence of a PEG concentration gradient induces an opposing diffusophoretic effect on the particle which reverses the sign of the Soret coefficient as shown in supplementary figure 2. This results in an effective Soret coefficient for the particles given by $S_T^* = S_T - 2\pi(S_T^{PEG} - \frac{1}{T})a\lambda^2 c$ [14], where $S_T$ is the intrinsic Soret coefficient of the particle, $S_T^{PEG}$ is the intrinsic Soret coefficient of the PEG molecules, $T$ is the local temperature and $\lambda$ is the interaction length scale for an entropic repulsion of the PEG molecules from the particle surface. We estimated the values of $S_T$ from the data provided in ref 18. Taking the values of $S_T^{PEG}$ and $\lambda$ to be $0.04\ K^{-1}$ and $5.2\ nm$ respectively as determined in previous works [17,20], we determined values for $S_T^*$ up to $-1.5\ K^{-1}$ for a $2.5\ mW$ laser power. We firstly investigated the relative effects of the optical and thermophoretic forces for a $100\ nm$ polystyrene bead. The optical force was computed using the Maxwell's Stress Tensor method while the thermophoretic force was computed according to the equation, $F_{therm} = K_B T(r) S_T \nabla T$, where $K_B$ represents the Boltzmann's constant. In figure 3(a) we show the relative strengths of the axial component of both forces for a $2.5\ mW$ trapping laser, varying the particle center along the Z axis from coordinates $(X = 0, Y = 0, Z = 85\ nm)$ to $(X = 0, Y = 0, Z = 385\ nm)$, where $Z = 30\ nm$ represents the top surface of the bowties. The repelling thermophoretic force exceeds the optical force causing particles to be repelled from the hotspot. Furthermore, the corresponding repulsive thermophoretic potential significantly surpasses the attractive optical potential as shown in figures 3(b) and (c). Next, we investigated the strength of the diffusophoretic force for varying laser powers as shown in figures 4(a-f). The in-plane force components were calculated at an axial distance $50\ nm$ from the top of the ITO film, while the axial component was calculated at X = Y = 0 along the Z axis. Our calculations show an axial force up to 1.6 times the value of the thermophoretic force shown in figure 3(a), ensuring that the particle is attracted towards the hot region. Furthermore, diffusophoretic potentials exceeding $-20\ K_B T$ is shown in figures 4(d-f) for $100 nm$ polystyrene beads. This ensures stable localization of the particle at the hotspot, overcoming the effects of Brownian motion. Next, we conducted experiments using the fabricated bowtie antennas in 10%, 12% and 14% PEG solutions.

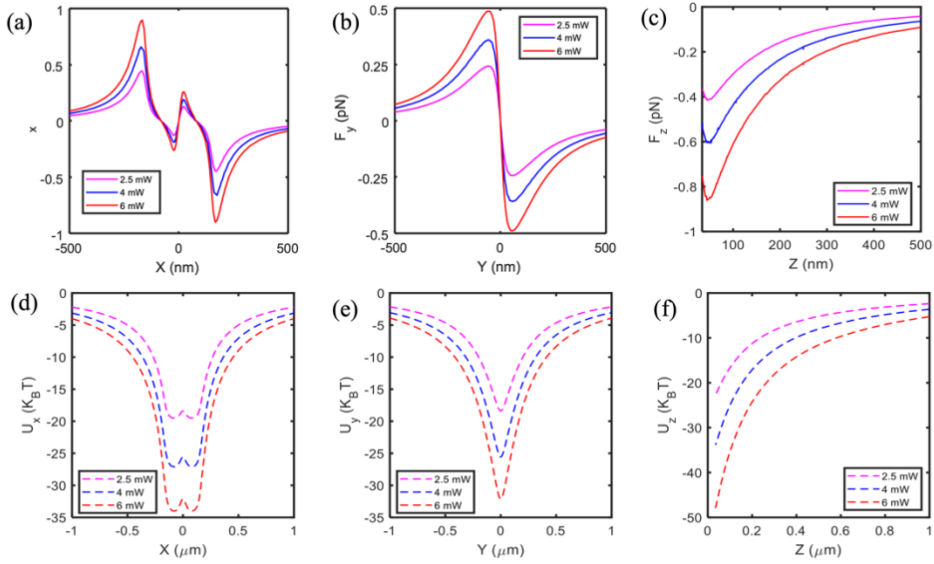

**Fig. 4.** Diffusophoretic and thermophoretic force calculations. (a) X-component diffusophoretic force along the X axis for a $100\ nm$ polystyrene bead, considering an X polarized $973\ nm$ laser illumination on the resonant bowtie antennas. The particle position is varied along the X axis while keeping the Z and Y positions at Z = $50\ nm$ and Y = 0. (b) Corresponding calculations for the Y component of the diffusophoretic force for the same illumination conditions as (a), while keeping the Z and X positions at Z = $50\ nm$ and X = 0. (c) Z component diffusophoretic force along the Z axis, keeping the in-plane coordinates at X = Y = 0 (center of the gap) and varying the Z coordinates. (d-f) show the corresponding diffusophoretic potentials for the forces shown in figures (a-c).

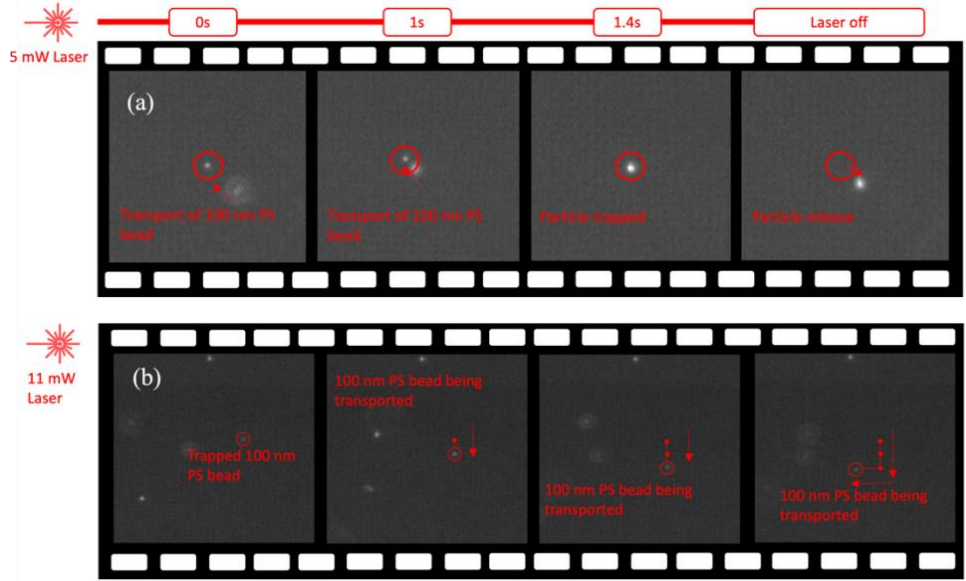

**Fig. 5.** Trapping and dynamic manipulation in a 10% PEG solution. (a) The figure captures the movement of a $100\ nm$ polystyrene bead towards the hotspot, where it is trapped using a $5\ mW$ laser. Subsequently, the particle is released after turning off the laser. (b) Dynamic 2D manipulation of a trapped particle is demonstrated under an $11\ mW$ laser illumination, following an "L" shaped trajectory depicted through frames 1 to 4.

Bowtie antennas were fabricated on an ITO coated glass substrate through a top-down nanofabrication approach. Firstly, bowtie patterns were created on a PMMA photoresist using electron beam lithography. Following the development of the patterned PMMA-coated substrate, Au was deposited onto the developed sample via electron beam evaporation, followed by a lift-off step. A detailed account of the fabrication process can be found in the supplementary information. To prevent the adsorption of negatively charged polystyrene beads on the bowtie antennas, the sample surface underwent treatment with poly (sodium 4-styrenesulfonate) (PSS), as previously described in our prior works [21,22]. A microfluidic chamber was assembled with $120\ \mu m$ dielectric spacers and a glass cover slip, effectively sealing the top of the chamber. For our experiments, we employed a $973\ nm$ diode laser focused to a spot size of $1.59\ \mu m$ (Airy disk) for trapping and a broadband fluorescent lamp with a $532\ nm$ filter for fluorescence excitation. Fluorescent images of particles were recorded using a CMOS camera (Photometrics PRIME 95B). All experiments were carried out on an inverted microscope (Nikon eclipse Ti2) with a $40 \times, 0.75\ NA$ objective lens.

100 nm Polystyrene beads as purchased from ThermoFisher Scientific were diluted by $10^6$ times in 10%, 12% and 14% PEG solutions and stored in a 4°C fridge prior to experiments.

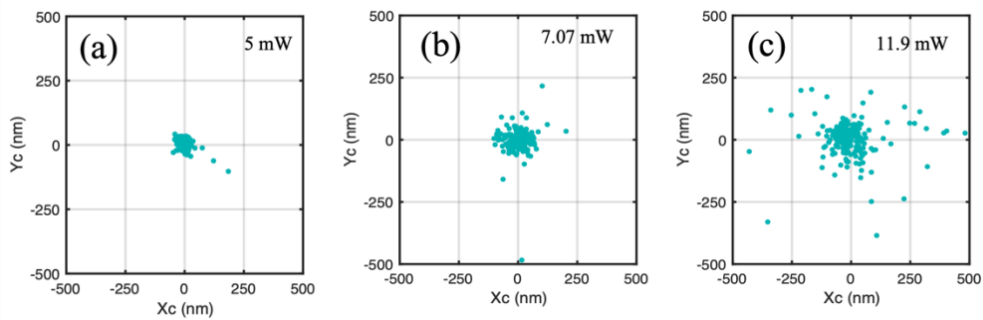

**Fig. 6.** Particle displacement scatter plots for 12% PEG solution. (a) Particle displacement scatter plots for a trapped 100 nm polystyrene bead at 5 $mW$ with a trap stiffness of 0.5 $fN/nm$, (b) 7.07 $mW$ with a stiffness of 0.25 $fN/nm$, and (c) 11.9 $mW$ with a stiffness of 0.09 $fN/nm$ This shows an apparent decease in the particle confinement with increasing laser power.

Firstly, we demonstrated the rapid transport and release of a 100 $nm$ polystyrene bead in a 10% PEG solution under a 5 $mW$ laser illumination as shown in Figure 5(a) (supplementary movie 1). Frames 1 and 2 show a single particle being transported, while frame 3 shows the particle trapped in place. Frame 4 shows the release of the particle from the trap. Using a bowtie array with $Px = Py = 1\ \mu m$, we demonstrated dynamic 2D manipulation of a trapped 100 $nm$ polystyrene bead as shown in figure 5(b) (supplementary movie 2). We showed the particle manipulated through what looks like an "L" shape in frames 1 through 4 using an 11 $mW$ laser power. This is made possible by the translation of the diffusophoretic potential from one antenna to the next due to the movement of the laser spot. Next, we investigated the dependence of the trap stability on the laser power. Trapping videos were obtained using a CMOS camera under a 100 $ms$ exposure time and analyzed using a custom Python script to obtain particle center displacement plots as shown in figures 6(a-c). The trap stiffnesses were calculated using the equipartition theorem and applying a motion blur correction function as detailed in the supplementary information. In Supplementary movies 3-12 we vary the laser power for a trapped particle in 10%, 12% and 14% PEG solutions. Figures 6(a-c) shows the tracked particle center displacement for a 100 $nm$ polystyrene bead in a 12% PEG solution for three different laser powers, obtained using our custom python script. A close examination of these plots reveals a tighter confinement of the particle with decrease in the laser power. Plotting the trap stiffness versus laser power for varying PEG concentrations as shown in supplementary figure 4 elucidates this trend (see Supplementary movies 4 - 11.). We obtain a trapping stiffness as high as 0.5 $fN/nm$ for a 5 $mW$ laser in a 12% PEG solution and a stiffness of approximately 0.3 $fN/nm$ for a 2.5 $mW$ in a 14% PEG solution (see supplementary video 8).

This observation can be explained by considering two counteracting effects. Firstly, it involves the interaction of neighboring diffusophoretic potentials resulting from the collective heating of the bowtie array, considering the periodic spacing. Neighboring antennas result in local potential wells that arise due to the overlap between the antennas and the laser spot, considering that the in-plane intensity distribution of the focused laser follows a Sinc-like function profile. We performed FEM thermal simulations considering the interaction of the adjacent antennas with the laser spot and show the presence of neighboring local diffusophoretic potentials, as depicted in supplementary figure 5. We also show that the ratio of the global potential depth to the local potentials decreases with increasing laser power. This could result in more a significant interaction of the trapped particle with local potentials at higher laser powers. However, our assessment of this effect might be an underestimation due to the uncertainties associated with precisely centering the laser spot around a given antenna. This uncertainty might lead to more significant overlap than accounted for in our simulations. The second effect stems from counteracting effects between thermophoresis and diffusophoresis, with thermophoresis resulting in an increase in the positive intrinsic Soret coefficient at higher temperatures. The overall effects of this results in a decrease in the magnitude of the effective negative Soret coefficient as shown in supplementary figure 2b. Previous works involving the thermophoretic depletion of DNA molecules in PEG solutions showed ring-like distributions due to the dominance of thermophoresis at the center of the hot region [13]. Furthermore, we demonstrate trapping for durations exceeding 1 minute in 12% and 14% PEG solutions, underscoring the efficacy of our platform for single particle localization and manipulation (see supplementary movie 3 and 12). Our results prove very effective for low power trapping of nanoscopic particles and show effective in-plane manipulation of trapped particles. This also shows great potential for 2D dynamic on-demand optical assembly of nanoscale particles.

## 3. Conclusion

The approach introduced in this work synergistically combines the depletion attraction and optical forces in the presence of thermally induced concentration gradients. The utilization of a depletion attraction force, resulting from thermophoretic depletion of polyethylene glycol (PEG) molecules, creates concentration gradients around the plasmonic hotspots. The diffusophoretic force generated by this concentration gradient efficiently localizes the nanoscopic particles, enabling stable trapping using less than $3\ mW$ input laser power. This presents opportunities for simultaneous stable trapping and *in situ* optical spectroscopy of trapped nanoscopic particles. Furthermore, we demonstrated dynamic 2D manipulation of $100\ nm$ polystyrene particles, showcasing the potential of this platform for colloidal assembly of nanoparticles. This work offers a promising solution for precise and efficient trapping and manipulation of nanoscopic particles, with potential implications in diverse fields such as biological and environmental sciences.

**Acknowledgements**

We acknowledge financial support from the National Science Foundation (NSF) CAREER Award (NSF ECCS 2143836).

**References**

1. Y. Pang and R. Gordon, "Optical trapping of a single protein," Nano Lett **12**, 402–406 (2012).

2. Y. Pang and R. Gordon, "Optical trapping of 12 nm dielectric spheres using double-nanoholes in a gold film," Nano Lett **11**, 3763–3767 (2011).

3. G. Volpe, O. M. Maragò, H. Rubinsztein-Dunlop, G. Pesce, A. B. Stilgoe, G. Volpe, G. Tkachenko, V. G. Truong, S. N. Chormaic, F. Kalantarifard, P. Elahi, M. Käll, A. Callegari, M. I. Marqués, A. A. R. Neves, W. L. Moreira, A. Fontes, C. L. Cesar, R. Saija, A. Saidi, P. Beck, J. S. Eismann, P. Banzer, T. F. D. Fernandes, F. Pedaci, W. P. Bowen, R. Vaippully, M. Lokesh, B. Roy, G. Thalhammer-Thurner, M. Ritsch-Marte, L. P. García, A. V. Arzola, I. P. Castillo, A. Argun, T. M. Muenker, B. E. Vos, T. Betz, I. Cristiani, P. Minzioni, P. J. Reece, F. Wang, D. McGloin, J. C. Ndukaife, R. Quidant, R. P. Roberts, C. Laplane, T. Volz, R. Gordon, D. Hanstorp, J. T. Marmolejo, G. D. Bruce, K. Dholakia, T. Li, O. Brzobohatý, S. H. Simpson, P. Zemánek, F. Ritort, Y. Roichman, V. Bobkova, R. Wittkowski, C. Denz, G. V. Pavan Kumar, A. Foti, M. G. Donato, P. G. Gucciardi, L. Gardini, G. Bianchi, A. V. Kashchuk, M. Capitanio, L. Paterson, P. H. Jones, K. Berg-Sørensen, Y. F. Barooji, L. B. Oddershede, P. Pouladian, D. Preece, C. B. Adiels, A. C. De Luca, A. Magazzù, D. B. Ciriza, M. A. Iatì, and G. A. Swartzlander, "Roadmap for optical tweezers," JPhys Photonics **5**, (2023).

4. J. C. Ndukaife, A. V. Kildishev, A. G. A. Nnanna, V. M. Shalaev, S. T. Wereley, and A. Boltasseva, "Long-range and rapid transport of individual nano-objects by a hybrid electrothermoplasmonic nanotweezer," Nat Nanotechnol **11**, 53–59 (2016).

5. W. Wei, X. Zhang, S. Zhang, G. Wei, and Z. Su, "Biomedical and bioactive engineered nanomaterials for targeted tumor photothermal therapy: A review," Materials Science and Engineering C **104**, (2019).

6.  A. C. V. Doughty, A. R. Hoover, E. Layton, C. K. Murray, E. W. Howard, and W. R. Chen, "Nanomaterial applications in photothermal therapy for cancer," Materials **12**, (2019).

7.  J. Gargiulo, T. Brick, I. L. Violi, F. C. Herrera, T. Shibanuma, P. Albella, F. G. Requejo, E. Cortés, S. A. Maier, and F. D. Stefani, "Understanding and Reducing Photothermal Forces for the Fabrication of Au Nanoparticle Dimers by Optical Printing," Nano Lett **17**, 5747–5755 (2017).

8.  B. J. Roxworthy, A. M. Bhuiya, S. P. Vanka, and K. C. Toussaint, "Understanding and controlling plasmon-induced convection," Nat Commun **5**, (2014).

9.  K. Namura, K. Nakajima, and M. Suzuki, "Investigation of transition from thermal- to solutal-Marangoni flow in dilute alcohol/water mixtures using nano-plasmonic heaters," Nanotechnology **29**, (2018).

10. Y. Liu, L. Lin, B. Bangalore Rajeeva, J. W. Jarrett, X. Li, X. Peng, P. Kollipara, K. Yao, D. Akinwande, A. K. Dunn, and Y. Zheng, "Nanoradiator-Mediated Deterministic Opto-Thermoelectric Manipulation," ACS Nano **12**, 10383–10392 (2018).

11. L. Lin, M. Wang, X. Peng, E. N. Lissek, Z. Mao, L. Scarabelli, E. Adkins, S. Coskun, H. E. Unalan, B. A. Korgel, L. M. Liz-Marzán, E. L. Florin, and Y. Zheng, "Opto-thermoelectric nanotweezers," Nat Photonics **12**, 195–201 (2018).

12. S. Yang, J. A. Allen, C. Hong, K. P. Arnold, S. M. Weiss, and J. C. Ndukaife, "Multiplexed Long-Range Electrohydrodynamic Transport and Nano-Optical Trapping with Cascaded Bowtie Photonic Crystal Nanobeams," Phys Rev Lett **130**, (2023).

13. Y. T. Maeda, T. Tlusty, and A. Libchaber, "Effects of long DNA folding and small RNA stem-loop in thermophoresis," Proc Natl Acad Sci U S A **109**, 17972–17977 (2012).

14. H. R. Jiang, H. Wada, N. Yoshinaga, and M. Sano, "Manipulation of colloids by a nonequilibrium depletion force in a temperature gradient," Phys Rev Lett **102**, (2009).

15. Y. T. Maeda, A. Buguin, and A. Libchaber, "Thermal separation: Interplay between the soret effect and entropic force gradient," Phys Rev Lett **107**, (2011).

16. M. He, J. Crow, M. Roth, Y. Zeng, and A. K. Godwin, "Integrated immunoisolation and protein analysis of circulating exosomes using microfluidic technology," Lab Chip **14**, 3773–3780 (2014).

17. J. Deng, F. Tian, C. Liu, Y. Liu, S. Zhao, T. Fu, J. Sun, and W. Tan, "Rapid One-Step Detection of Viral Particles Using an Aptamer-Based Thermophoretic Assay," J Am Chem Soc **143**, 7261–7266 (2021).

18. M. Braibanti, D. Vigolo, and R. Piazza, "Does thermophoretic mobility depend on particle size?," Phys Rev Lett **100**, (2008).

19. D. G. Kotsifaki and S. Nic Chormaic, "The role of temperature-induced effects generated by plasmonic nanostructures on particle delivery and manipulation: A review," Nanophotonics **11**, 2199–2218 (2022).

20. C. Liu, J. Zhao, F. Tian, L. Cai, W. Zhang, Q. Feng, J. Chang, F. Wan, Y. Yang, B. Dai, Y. Cong, B. Ding, J. Sun, and W. Tan, "Low-cost thermophoretic profiling of extracellular-vesicle surface proteins for the early detection and classification of cancers," Nat Biomed Eng **3**, 183–193 (2019).

21. T. Anyika, C. Hong, and J. C. Ndukaife, "High-speed nanoscale optical trapping with plasmonic double nanohole aperture," Nanoscale (2023).

22. C. Hong, S. Yang, and J. C. Ndukaife, "Stand-off trapping and manipulation of sub-10 nm objects and biomolecules using opto-thermo-electrohydrodynamic tweezers," Nat Nanotechnol **15**, 908–913 (2020).

# Supplementary Information

**Thermal simulations**

The heat source density generated due to the electric field around the plasmonic antennas is given as $q(r) = 1/Re(\boldsymbol{J} \cdot \boldsymbol{E})$. Where $\boldsymbol{J}$ is the current density induced in the vicinity of the plasmonic aperture. The temperature rise for the system was computed using COMSOL Multiphysics by solving the heat transfer equation given by,

$$-k\nabla^2 T(r) + \rho c_p \boldsymbol{u} \cdot \nabla T(r) = q(r) \quad (1).$$

Where $k$ is the thermal conductivity, $T(r)$ is the temperature field, $\rho$ the fluid density, and $c_p$ is the specific heat capacity.

**Supplementary figure 1**

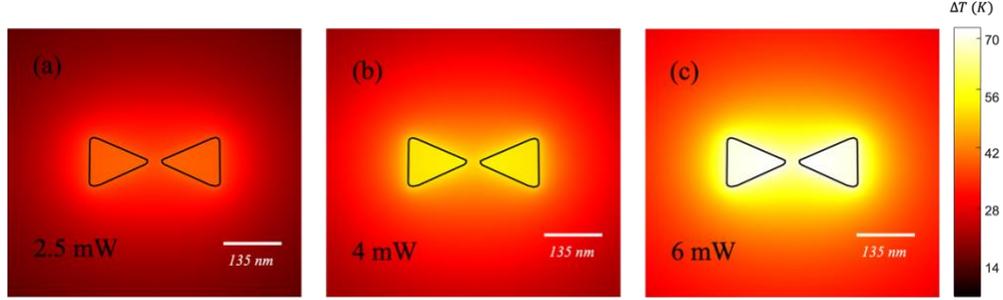

**S1.** Thermal simulations for three different laser powers showing significant temperature rise for this system.

**PEG depletion simulation**

The transport of PEG molecules in the presence of temperature gradients is governed by the flux equation,

$$j_{PEG} = cS_T^{PEG} D_{PEG}\nabla T - \nabla c D_{PEG} \qquad (2).$$

Where $j_{PEG}$ is the flux, $c$ is the PEG concentration, $D_{PEG}$ is the Brownian diffusion coefficient of the PEG molecules, $S_T^{PEG}$ is the Soret coefficient of the PEG molecules and $\nabla T$ is the local temperature gradient. To obtain the steady state PEG concentration profile, we solved the transport equation given by,

$$\frac{\partial c}{\partial t} = \nabla c D_{PEG} - cS_T^{PEG} D_{PEG}\nabla T \qquad (3).$$

Where $S_T^{PEG} = 0.04 K^{-1}$ [17] and $D_{PEG} = 9.3 \times 10^{-11} m^2/s$ [17]. The Diffusophoretic force on $100\ nm$ polystyrene particles was computed from the equation $F_d = -3\pi\eta a D S_T^* \nabla T$, where $a$ is the particle diameter, $\eta$ is the viscosity of the solution, $D$ is the Brownian diffusion coefficient of the polystyrene bead, $S_T^*$ is the effective soret coefficient of the particle which is proportional to the PEG concentration $c$. The value of the intrinsic soret coefficient $S_T$ estimated using the data provided in figure 1 of reference 18 as shown in S2. The curve for $S_T$ was plotted according to the empirical equation,

$$S_T = S_T^\infty \left[1 - e^{\frac{T^*-T}{T_0}}\right] \tag{4}$$

Where $S_T^\infty$ is the high temperature limit of the soret coefficient taken to be $1.93\ K^{-1}$, $T^*$ is the temperature at which the value of the soret coefficient changes its sign, and $T_0$ is taken to be $0.1T$. The effective soret coefficient for particles in the depleted region is governed by the relative strengths of diffusophoresis and thermophoresis and is given by the equation,

$$S_T^* = S_T - 2\pi(S_T^{PEG} - \frac{1}{T})a\lambda^2 c \tag{5}$$

Where $\lambda$ is the interaction length scale for an entropic repulsion of the PEG molecules from the particle surface and is taken to be $5.2\ nm$.

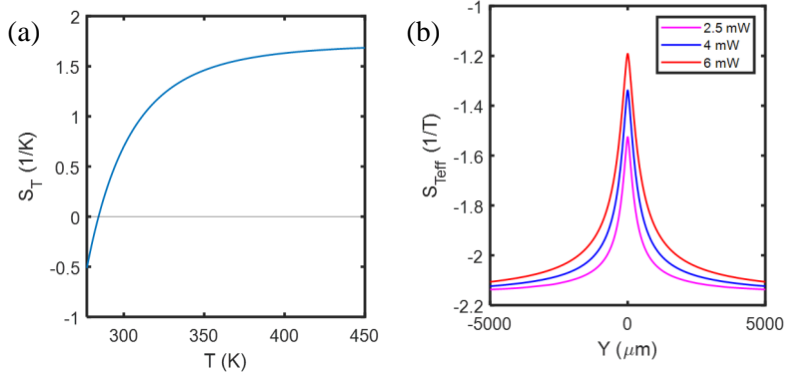

S2. (a) Soret coefficient for 100 nm polystyrene beads estimated from reference 18, according to equation 4. (b) Spatial variation of the effective soret coefficient for various laser powers

## Sample Fabrication

Fused silica substrate coated with 50 nm ITO later were washed with acetone then rinsed with IPA, followed by a 10 second oxygen plasma cleaning (Trion Phantom II) process. Afterwards, PMMA A4 was spin-coated on the samples and subsequently patterned using electron beam lithography. Subsequently the samples were developed in MIBK-IPA for 35 seconds. A $30\ nm$ thick gold film was then deposited on the developed samples in a multimode deposition chamber (Angstrom Amod – Combined e-beam, Resistive & Sputter Deposition chamber). Lastly, PMMA was removed using 1165 (n-methyl pyrrolidinone). To prevent adsorption of negatively charged polystyrene beads to the Au film, we treated the Au surface with negatively charged Poly (sodium 4-styrenesulfonate) (PSS) for 10 minutes and rinsed with potassium chloride (KCl) for 5 minutes followed by deionized water. Next, $120\ \mu m$ dielectric spacers were used to define the chamber and a glass cover slip was used to cover the chamber.

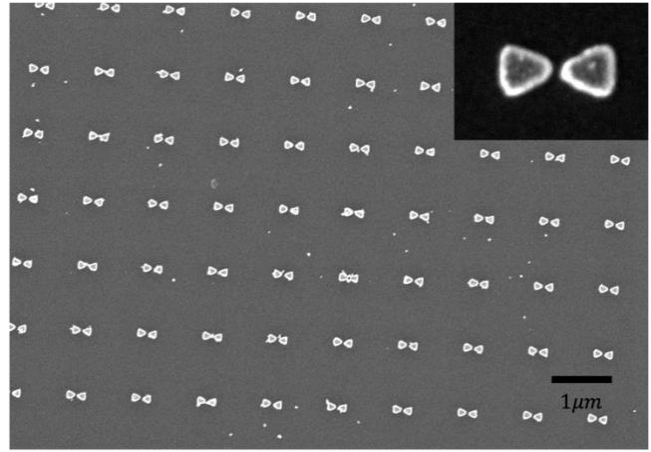

**S3.** Scanning electron micrograph of the fabricated sample.

**Determination of trap stiffness**

The finite integration time of our acquired videos results in a motion blur effect which results in an underestimation of the trap stiffness. To correct for this, we applied a motion blur correction function defined as,

$$Var_m(i) = \frac{K_B T}{k_i} S(\alpha) \tag{6}$$

Where $Var_m(i)$ is the experimentally measured variance, $S(\alpha)$ is the motion blur correction function, $K_B$ is the Boltzmann constant, and $k_i$ is the stiffness along the $ith$ axis. The trap stiffness $k_i$ was obtained by numerically solving equation 6, using the exact form of $S(\alpha)$ as defined in reference [1] of this SI.

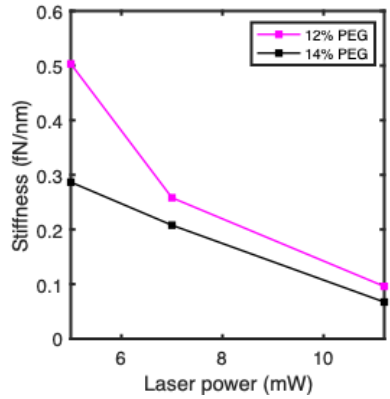

**S4.** Trap stiffness versus laser power for $100\ nm$ polystyrene beads in 12% and 14% PEG solutions

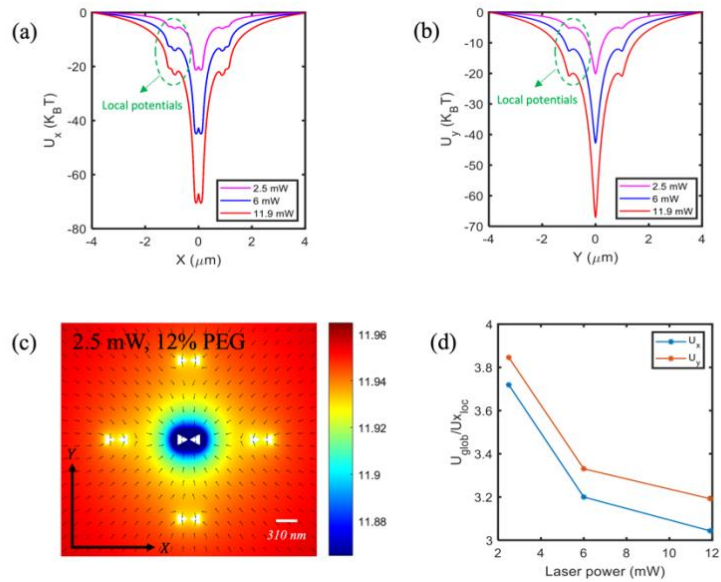

**S5.** Diffusophoretic potential simulations for 1 micron array bowtie antennas. (a) Diffusophoretic potential along the X axis for a 100 $nm$ polystyrene bead, considering an X polarized 973 $nm$ laser illumination on the resonant bowtie antennas for different laser powers. The particle position is varied along the X axis while keeping the Z and Y positions at Z = 50 $nm$ and Y = 0. (c) Corresponding diffusophoretic potential along the Y axis. The dotted curves in (a) and (b) show local potentials that arise due to the overlap of neighboring antennas with the laser. (c) PEG concentration profile for a 2.5 $mW$ laser illumination in a 12% PEG solution. (d) The ratio of the global potential depth to the local potential decreases with increasing laser power, suggesting a more significant interaction of the particle with these local potentials.

**References**

1.     W. P. Wong and K. Halvorsen, "The effect of integration time on fluctuation measurements: calibrating an optical trap in the presence of motion blur," Opt. Express **16**, 12517–12531 (2006).

15